\documentclass[pra,preprint,amssymb]{revtex4-1}
\usepackage{enumerate,amsmath}
\usepackage{graphicx}
\usepackage{color}

\begin{document}

\title{Hardness of classically sampling
one clean qubit model with constant total variation distance error}
\author{Tomoyuki Morimae}
\email{morimae@gunma-u.ac.jp}
\affiliation{Department of Computer Science, 
Gunma University, 1-5-1 Tenjincho, Kiryu,
Gunma, 376-0052, Japan}

\date{\today}
\begin{abstract}
The one clean qubit model (or the DQC1 model) is a restricted model of quantum
computing where only a single input qubit
is pure and all other input qubits are maximally mixed.
In spite of the severe restriction,
the model can solve several problems (such as calculating Jones
polynomials) whose classical efficient solutions are not known.
Furthermore, it was shown that if the output probability distribution
of the one clean qubit model can be classically efficiently sampled
with a constant multiplicative error, then the polynomial hierarchy collapses
to the second level.
Is it possible to improve the multiplicative error hardness
result to a constant total variation distance error one
like other
sub-universal quantum computing models such as the IQP model,
the Boson Sampling model, and the Fourier Sampling model?
In this paper, we show that
it is indeed possible if we accept a modified
version of the average case hardness conjecture.
Interestingly, the anti-concentration lemma can be easily shown by
using the special property of the one clean qubit model
that each output probability is so small that
no concentration occurs.
\end{abstract}

\maketitle

\section{Introduction}
The one clean qubit model (or the DQC1 model)
first introduced by Knill and Laflamme~\cite{KnillLaflamme}
is a restricted model
of quantum computing where only a single input qubit
is pure and all other input qubits are maximally mixed. 
In spite of the severe restriction,
surprisingly, 
the model can solve several problems whose efficient classical solutions 
are not known, such as
the spectral density estimation~\cite{KnillLaflamme}, 
testing integrability~\cite{integrability}, calculations of 
the fidelity decay~\cite{fidelity_decay}, and approximations of the Jones 
polynomial, HOMFLY polynomial, and Turaev-Viro 
invariant~\cite{ShorJordan08,Passante09,JordanWocjan09,JordanAlagic}.
Furthermore, it was recently shown that if the output probability distribution
of the one clean qubit model is classically efficiently sampled
with a constant multiplicative error, then the polynomial hierarchy
collapses to the second level~\cite{MFF,FKMNTT}.
(Here, we say that a probability distribution 
$\{p_z\}_z$ is sampled by a machine $M$ 
with a multiplicative error $\epsilon\ge0$ if
\begin{eqnarray*}
|p_z-q_z|\le \epsilon p_z
\end{eqnarray*}
is satisfied for all $z$,
where $\{q_z\}_z$ is the output probability distribution
of $M$.)
Since a collapse of the polynomial hierarchy is not believed to happen
in computer science, the result suggests the impossibility
of classically simulating the one clean qubit model.
Similar hardness results for constant multiplicative error sampling were also 
shown for other sub-universal quantum computing
models, such as the IQP model~\cite{BJS} 
and the Boson Sampling model~\cite{AA}.

The requirement of constant multiplicative error sampling is, however, strong,
and sampling with a constant 
total variation distance error (or the L1-norm error) is 
considered as more appropriate.
(Here, we say that a probability distribution $\{p_z\}_z$ is
sampled by a machine $M$ 
with a total variation distance error $\epsilon\ge0$ if
\begin{eqnarray*}
\sum_z|p_z-q_z|\le\epsilon
\end{eqnarray*}
is satisfied, where $\{q_z\}_z$ is the output probability distribution
of $M$.)
In fact, the hardness results with constant total variation distance errors
were shown for
the IQP model~\cite{BMS}, 
the Boson Sampling model~\cite{AA},
and the Fourier Sampling model~\cite{FU}
(assuming some conjectures).
Is it possible to show a similar
constant-total-variation-distance-error
hardness result for the one clean qubit model?

In this paper, we 
show that it is indeed possible if we accept a modified
version of the average case hardness conjecture. 
Our proof is similar to those of Refs.~\cite{AA,BMS,FU},
but there is one interesting difference
which is specific to the one clean qubit model:
the anti-concentration lemma can be easily shown.
For the Boson Sampling model and the Fourier Sampling model, 
the anti-concentration lemma is a 
conjecture~\cite{AA,FU}. 
For the IQP model, it is shown with some calculations by using a special
structure of IQP circuits~\cite{BMS}. For the present case,
as we will see later, 
the anti-concentration lemma is easily shown by using 
a special property of the one clean qubit model that 
each probability is so small that no concentration occurs.

\section{Average case hardness conjecture}
As in the cases of other sub-universal quantum computing
models, such as the IQP model~\cite{BMS}, 
the Boson Sampling model~\cite{AA}, and the Fourier Sampling
model~\cite{FU}, we need a conjecture 
so-called ``average case hardness conjecture",
which claims that the $\#$P-hardness for the worst case
can be lifted to an average case. 
To show our result, which is a hardness of efficient classical sampling
of the one clean qubit model with a constant total variation
distance error, we need the following conjecture.

{\bf Conjecture}:
For each $n$, there exists a discrete set
${\mathcal U}_{n+1}$ of uniformly-generated polynomial-time
$(n+1)$-qubit unitary operators
such that
calculating
\begin{eqnarray*}
f(z,U)\equiv
\langle z|U(|0\rangle\langle0|\otimes I^{\otimes n})U^\dagger|z\rangle
\end{eqnarray*}
with a multiplicative error less than 1/2 for
more than 1/6 fraction of $(z,U)\in\{0,1\}^{n+1}\times{\mathcal U}_{n+1}$
is $\#$P-hard.

Here, $I\equiv|0\rangle\langle0|+|1\rangle\langle1|$
is the two-dimensional identity operator, and
$|z\rangle$ is the computational-basis state corresponding to
the bit string $z$.
(For example, if $z=010$, 
$|z\rangle=|010\rangle=|0\rangle\otimes|1\rangle\otimes|0\rangle$.)

Unfortunately,
we do not know whether the conjecture is true or not, but
we can show that it is true at least for the worst case.
Here, we give two proofs for the worst-case $\#$P-hardness.

\subsection{First proof}
Let us consider a unitary operator $U$ such that
\begin{eqnarray*}
U^\dagger&=&\Big[I\otimes|0\rangle\langle0|^{\otimes n}
+X\otimes(I^{\otimes n}-|0\rangle\langle0|^{\otimes n})\Big]
(I\otimes C),
\end{eqnarray*}
where $C$ is an $n$-qubit IQP circuit.
Then,
\begin{eqnarray*}
f(0^{n+1},U)&=&|\langle0^n|C|0^n\rangle|^2.
\end{eqnarray*}
For certain IQP circuits $C$,
$\langle0^n|C|0^n\rangle$
is related to
the partition function, $Z$, of the Ising model~\cite{BMS,FujiiMorimae} and 
the gap function, $gap(f)$, of a 
degree-3 polynomial $f$ over ${\mathbb F}_2$~\cite{BMS}:
\begin{eqnarray*}
\langle0^n|C|0^n\rangle&=&\frac{Z}{2^n},\\
\langle0^n|C|0^n\rangle&=&\frac{gap(f)}{2^n}.
\end{eqnarray*}
It is known that calculating $|Z|^2$ and $gap(f)^2$
with constant multiplicative errors is $\#$P-hard~\cite{BMS,FujiiMorimae}. 
Hence calculating 
$|\langle0^n|C|0^n\rangle|^2=f(0^{n+1},U)$
with constant multiplicative errors 
is $\#$P-hard for certain unitary operators $U$,
which shows the correctness of the conjecture
for the worst case. 

\subsection{Second proof}
We define two unitary operators
$U_1$ and $U_2$ as
\begin{eqnarray*}
U_1^\dagger&=& 
\Big[
(I\otimes|0\rangle\langle0|+X\otimes|1\rangle\langle1|)
\otimes I^{\otimes n-1}\Big]
(I\otimes V),\\
U_2^\dagger&=& 
\Big[
\Big(
I\otimes|0\rangle\langle0|^{\otimes 2}
+X\otimes(I^{\otimes 2}-|0\rangle\langle0|^{\otimes 2})
\Big)
\otimes I^{\otimes n-2}\Big](I\otimes V),
\end{eqnarray*}
where $V$ is an $n$-qubit unitary operator.
Then, we obtain
\begin{eqnarray*}
f(0^{n+1},U_1)&=&\langle0^n|V^\dagger(|0\rangle\langle0|
\otimes I^{\otimes n-1})V|0^n\rangle,\\  
f(0^{n+1},U_2)&=&\langle0^n|V^\dagger(|0\rangle\langle0|^{\otimes 2}
\otimes I^{\otimes n-2})V|0^n\rangle.
\end{eqnarray*}
Now we show that calculating 
$f(0^{n+1},U_1)$ and $f(0^{n+1},U_2)$ with
a constant multiplicative error $0\le \epsilon<1$ is
postBQP-hard. 
Since ${\rm postBQP}={\rm PP}$~\cite{postBQP}
and ${\rm P}^{\rm PP}={\rm P}^{\#{\rm P}}$,
it means that the calculation is $\#$P-hard.
Proof is as follows. Let us assume that there exists 
an algorithm that calculates $a$ and $b$ such that
\begin{eqnarray*}
|f(0^{n+1},U_1)-a|&\le& \epsilon f(0^{n+1},U_1),\\
|f(0^{n+1},U_2)-b|&\le& \epsilon f(0^{n+1},U_2).
\end{eqnarray*}
Let $L$ be a language in postBQP. 
Then, for any polynomial $r$, there exists a uniform
family $\{V_x\}_x$ of polynomial-time quantum circuits such that
\begin{itemize}
\item
If $x\in L$ then $P_{V_x}(o=0|p=0)\ge1-2^{-r}$.
\item
If $x\notin L$ then $P_{V_x}(o=0|p=0)\le2^{-r}$.
\end{itemize}
Here,
\begin{eqnarray*}
P_{V_x}(o=0|p=0)=\frac{P_{V_x}(o=0,p=0)}{P_{V_x}(p=0)},
\end{eqnarray*}
$P_{V_x}(o=0,p=0)$ is the probability that $V_x$ outputs
$(o,p)=(0,0)$,
and
$P_{V_x}(p=0)$ is the probability that $V_x$ outputs
$p=0$.

Let us construct $U_1$ and $U_2$ by using $V_x$.
Then, for any polynomial $r$, if $x\in L$,
\begin{eqnarray*}
\frac{b}{a}
&\ge&
\frac{1-\epsilon}{1+\epsilon}
\frac{f(0^{n+1},U_2)}
{f(0^{n+1},U_1)}\\
&\ge&
\frac{1-\epsilon}{1+\epsilon}
(1-2^{-r}),
\end{eqnarray*}
and
if $x\notin L$,
\begin{eqnarray*}
\frac{b}{a}
&\le&
\frac{1+\epsilon}{1-\epsilon}
\frac{f(0^{n+1},U_2)}
{f(0^{n+1},U_1)}\\
&\le&
\frac{1+\epsilon}{1-\epsilon}
2^{-r}.
\end{eqnarray*}
Therefore, if we can calculate $a$ and $b$, we can solve $L$.

\section{Main result}
If we accept the conjecture,
we can show the following theorem, which is the main result
of the present paper.

{\bf Theorem}:
If there exists a probabilistic polynomial-time
classical
algorithm that 
outputs $z\in\{0,1\}^{n+1}$ with probability $q_z(U)$
such that
\begin{eqnarray*}
\sum_{z\in\{0,1\}^{n+1}}|p_z(U)-q_z(U)|\le\epsilon
\end{eqnarray*}
for any $U\in\cup_n{\mathcal U}_{n+1}$,
then the polynomial hierarchy collapses to the third level.
Here, $\epsilon=\frac{1}{36}$
and
\begin{eqnarray*}
p_z(U)\equiv\langle z|U\Big(|0\rangle\langle0|\otimes
\frac{I^{\otimes n}}{2^n}\Big)
U^\dagger|z\rangle.
\end{eqnarray*}

The theorem says that if the output probability distribution $p_z(U)$
of the one clean qubit model can be classically efficiently sampled
with the total variation distance error $\epsilon$,
then the polynomial hierarchy collapses to the third level.

{\bf Proof}:
Now let us give a proof of the theorem.
Our proof is similar to those of Refs.~\cite{AA,BMS,FU}
except that the anti-concentration lemma can be easily shown.

Let $\delta>0$ be a parameter specified later.
From Markov's inequality,
\begin{eqnarray*}
\mbox{Pr}_{z,U}
\Big[|p_z(U)-q_z(U)|\ge
\frac{\epsilon}{2^{n+1}\delta}\Big]
&\le&\frac{2^{n+1}\delta}{\epsilon}
\frac{1}{2^{n+1}|{\mathcal U}_{n+1}|}
\sum_{U,z}|p_z(U)-q_z(U)|\\
&\le&\delta.
\end{eqnarray*}
From Stockmeyer's Counting Theorem~\cite{Stockmeyer}, 
there exists an ${\rm FBPP}^{\rm NP}$
algorithm that outputs $\tilde{q}_z(U)$ such that
\begin{eqnarray*}
|\tilde{q}_z(U)-q_z(U)|\le\frac{q_z(U)}{poly}.
\end{eqnarray*}
Therefore,
\begin{eqnarray*}
|\tilde{q}_z(U)-p_z(U)|&\le&|\tilde{q}_z(U)-q_z(U)|+|q_z(U)-p_z(U)|\\
&\le&\frac{q_z(U)}{poly}+|q_z(U)-p_z(U)|\\
&=&\frac{p_z(U)+q_z(U)-p_z(U)}{poly}
+|q_z(U)-p_z(U)|\\
&\le&\frac{p_z(U)+|q_z(U)-p_z(U)|}{poly}
+|q_z(U)-p_z(U)|\\
&=&\frac{p_z(U)}{poly}+|q_z(U)-p_z(U)|\Big(1+\frac{1}{poly}\Big)\\
&<&\frac{p_z(U)}{poly}
+\frac{\epsilon}{2^{n+1}\delta}\Big(1+\frac{1}{poly}\Big)
\end{eqnarray*}
with more than $1-\delta$ fraction of $(z,U)$.

Let $S\subseteq\{0,1\}^{n+1}\times {\mathcal U}_{n+1}$ 
be the set of $(z,U)$ such that
\begin{eqnarray*}
\frac{\epsilon}{2^{n+1}\delta}\le\frac{p_z(U)}{3}.
\end{eqnarray*}
Since 
\begin{eqnarray*}
p_z(U)&=&\langle z|U\Big(|0\rangle\langle0|\otimes\frac{I^{\otimes n}}{2^n}
\Big)U^\dagger|z\rangle\\
&=&\frac{1}{2^n}
\langle z|U(|0\rangle\langle0|\otimes I^{\otimes n}
)U^\dagger|z\rangle\\
&\le&\frac{1}{2^n}\times1=\frac{1}{2^n},
\end{eqnarray*}
for all $(z,U)$,
and
\begin{eqnarray*}
\sum_{z\in\{0,1\}^{n+1}}p_z(U)=1
\end{eqnarray*}
for all $U$,
we obtain
\begin{eqnarray*}
1&=&\frac{1}{|{\mathcal U}_{n+1}|}
\sum_{U,z}p_z(U)\\
&=&\frac{1}{|{\mathcal U}_{n+1}|}
\sum_{(z,U)\in S}p_z(U)
+\frac{1}{|{\mathcal U}_{n+1}|}\sum_{(z,U)\notin S}p_z(U)\\
&<&\frac{1}{2^n|{\mathcal U}_{n+1}|}|S|
+\frac{2^{n+1}|{\mathcal U}_{n+1}|-|S|}{|{\mathcal U}_{n+1}|}
\frac{3\epsilon}{2^{n+1}\delta},
\end{eqnarray*}
which means
\begin{eqnarray*}
\frac{|S|}{2^{n+1}|{\mathcal U}_{n+1}|}&>&
\frac{1-\frac{3\epsilon}{\delta}}
{2-\frac{3\epsilon}{\delta}}.
\end{eqnarray*}

Therefore,
\begin{eqnarray*}
|\tilde{q}_z(U)-p_z(U)|
&<&\frac{p_z(U)}{poly}+\frac{p_z(U)}{3}\Big(1+\frac{1}{poly}\Big)\\
&=&p_z(U)\Big(\frac{1}{3}+\frac{1}{poly}\Big)
\end{eqnarray*}
for more than 
\begin{eqnarray*}
F\equiv1-\delta-\frac{1}{2-\frac{3\epsilon}{\delta}}
\end{eqnarray*}
fraction of $(z,U)$.
For example, if we take
$\delta=6\epsilon$, 
\begin{eqnarray*}
F=\frac{1}{6}.
\end{eqnarray*}

Note that
\begin{eqnarray*}
p_z(U)=\frac{f(z,U)}{2^n}.
\end{eqnarray*}
Therefore, the above result means that
there exists an ${\rm FBPP}^{\rm NP}$ algorithm that outputs
$\tilde{q}_z(U)$ such that
\begin{eqnarray*}
|\tilde{q}_z(U)2^n-f(z,U)|
&<&f(z,U)\Big(\frac{1}{3}+\frac{1}{poly}\Big)
<\frac{1}{2}f(z,U)
\end{eqnarray*}
for more than 
1/6 fraction of $(z,U)$.
If our average-case hardness conjecture is true,
it means the collapse of the polynomial hierarchy to the third level.

\acknowledgements
TM is supported by JST ACT-I No.JPMJPR16UP,
the Grant-in-Aid for Scientific Research on Innovative Areas
No.15H00850 of MEXT Japan, 
and the JSPS Grant-in-Aid for Young Scientists (B) 
No.26730003 and No.17K12637.

\end{document}